\documentstyle{article}
\newcommand{\beginabstract}{\begin{abstract}}
\def\scri{\hbox{${\cal J}$\kern -.645em {\raise
      .57ex\hbox{$\scriptscriptstyle (\ $}}}}
\newcommand{\mhat}{{\check M}} \newcommand{\Ximap}{{\Psi}}
\newcommand{\mzero}{{\tilde M}} \newcommand{\gone}{g^1}
\newcommand{\cD}{{\cal{D}}} \newcommand{\wnetrze}[1]{\hat{#1}}
\newcommand{\cO}{{\cal{O}}} \newcommand{\cH}{{\cal{H}}}
\newcommand{\cK}{{\cal K}}

 \newcommand{\cB}{{\cal B}}
\newcommand{\cC}{{\cal C}} \newcommand{\cV}{{\cal V}}
 \newcommand{\prt}{\partial}

 \newcommand{\eq}[1]{(\ref{#1})}
 \newcommand{\cU}{{\cal U}}

\newcommand{\commentout}[1]{}  
 
 \newcommand{\be}{\begin{equation}}

\newcommand{\ee}{\end{equation}} \newcommand{\bea}{\begin{eqnarray}}
\newcommand{\eea}{\end{eqnarray}}
\newcommand{\beaa}{\begin{eqnarray*}}
\newcommand{\eeaa}{\end{eqnarray*}}

\newtheorem{Theorem}   {Theorem}   [section]

\newtheorem{Lemma}     [Theorem]   {Lemma}
\newtheorem{Proposition} [Theorem] {Proposition}

 \newcommand{\R}{I\!\! R}

\newcommand{\ben}{\begin{enumerate}}
\newcommand{\een}{\end{enumerate}} \newcommand{\DA}{\mbox{\scriptsize
    DA}}
\title{On the dynamics of generators of Cauchy horizons}
\author{P.\  T.\ Chru\'sciel\thanks{Alexander von Humboldt fellow, on
    leave of absence from the Institute of Mathematics, Polish Academy
    of Sciences, Warsaw. e-mail: piotr@mpa-garching.mpg.de} \\
Max Planck Institut f\"ur Astrophysik\\ D 85740 Garching
  bei M\"unchen, GERMANY \\ \\  J.
  Isenberg\thanks{e-mail: jim@newton.uoregon.edu}\\
 Institute for Theoretical
  Science\\ University of Oregon\\ Eugene, OR 97403, USA
}
\begin{document}
\maketitle
%
\beginabstract
We discuss various features of the dynamical system determined by the
flow of null geodesic generators of Cauchy horizons. Several examples
with non--trivial (``chaotic'', ``strange attractors'', etc.) global
behaviour are constructed. Those examples  are relevant to
the ``chronology protection conjecture'', and they show that the
occurrence of ``fountains'' is {\em not} a generic feature of Cauchy
horizons.
\end{abstract}

\section{Introduction}
\label{Section 1}
In considering the question of whether the laws of physics prevent one
from constructing a ``time machine", Hawking \cite{Hawking} and Thorne
\cite{Thorne} have both stressed the importance of understanding the
generic behavior of the null generators of compactly generated Cauchy
horizons.  In particular it has been suggested ({\em cf.\ e.g.\/}
\cite{Thorne,Visser} and references therein) that the onset of quantum
instabilities in Cauchy horizons containing ``fountains'' would
prevent the formation of time machines. Here a ``fountain'' on a
future Cauchy horizon is defined as a periodic\footnote{Throughout
  this paper ``periodic" means periodic as a path in spacetime (and
  {\em not} necessarily as a path in the tangent bundle).} generator
$\gamma$ of the horizon such that a ``nonzero--measure'' set of
generators of the horizon asymptotically approaches $\gamma$ when
followed backwards in time.  It is therefore of some interest to
enquire whether or not the existence of fountains is a generic
property of ``compactly generated'' Cauchy horizons. In this work we
wish to point out that this is unlikely to
be true: we construct 
spacetimes with
compactly generated  Cauchy horizons for which no fountains occur.

When discussing features of Cauchy horizons, one should focus on
features which are stable in an appropriate sense.  We show that in
the set of all spacetimes with compactly generated Cauchy horizons,
there are open sets consisting entirely of spacetimes with
nonfountain--like behavior.  Unfortunately we are able to make
rigorous claims only for compact Cauchy horizons.  So the possibility
remains open that for spacetimes with compactly generated Cauchy
horizons which are {\it not} compact, fountains could generically
occur. While it is clear to us that this is not true, we note that
there is an important technical difference between compact Cauchy
horizons and noncompact yet compactly generated Cauchy horizons: As we
show in Section \ref{Section 4}, if a Cauchy horizon is compactly
generated and noncompact, and if further it is contained in an
asymptotically flat spacetime (in a technical sense made precise in
that Section), then the generators of the Cauchy horizon cannot be
continuous.  This is one of the difficulties which one has to face
when trying to make any rigorous claims about the dynamics of the
generators of some non--compact Cauchy horizons.

It is important to note that, following \cite{Hawking,Thorne}, we do
not impose any field equations on the spacetimes under
consideration.
Recall that one expects the existence of a  Cauchy horizon
to be an unstable feature, when the Einstein field equations (vacuum, or
with energy conditions on the source fields) are imposed. It would be
interesting to carefully investigate the extent to which the imposition of
field equations restricts the allowed dynamics of Cauchy horizon generators;
however this 
problem is not addressed here.

After discussing some preliminary definitions and ideas
in Section \ref{Section 2}, we focus on verifying the existence of the
spacetimes with
nonfountain-like dynamics, first for compact Cauchy horizons (Section
\ref{Section 3}) and then
for compactly generated but noncompact Cauchy horizons (Section
\ref{Section 4}).  The discussion
of noncompact Cauchy horizons in Section \ref{Section 4} includes the proof
that if the spacetime
containing it is asymptotically flat, then the generators cannot be
continuous.  

\section{Preliminaries on Cauhy horizons and dynamical systems}
\label{Section 2}
We shall consider $C^k$, $ (k \geq 3)$ spacetimes $(M^4, g)$ which
contain Cauchy horizons (we use the terminology of \cite{HE}).
Standard results \cite{HE} show that a Cauchy horizon is foliated by a
congruence of null geodesics.  These are called the {\it generators}
of the horizon.  One finds that if one follows a generator of a future
Cauchy horizon into its past then the generator always remains inside
the horizon.  This is not necessarily true if one follows a generator
(on a future Cauchy horizon) into its future.  We shall say that a
future Cauchy horizon ${\cal H}^+$ is {\em compactly generated}, if
there exists a compact set ${\cal K}\subset M$ such that every
generator of ${\cal H}^+$ enters and remains in ${\cal K}$, when
followed into the past.

To discuss the behavior of the generators of a Cauchy horizon, we wish
to use some of the language of dynamical systems theory.  Recall that
a
dynamical system $(\Sigma^n,X)$ consists of an 
$n$--dimensional manifold $\Sigma^n$ and a vector field $X$
specified on $\Sigma^n$.
Note that a 
Cauchy horizon ${\cal H}$  together with the vector field $
T$ of 
tangents
to its  generators (normalized in an arbitrary way) constitutes a
dynamical system $({\cal H} , T)$. We shall always choose  the
{\em past directed}
orientation of the generators  on a {\em future} Cauchy horizon.
For future Cauchy horizons the past--oriented
generators of ${\cal H}$  are then the orbits of this dynamical system.  A
distinguished feature of a Cauchy horizon when viewed as a dynamical system is
that the vector field $T$ is nowhere vanishing, so none of the orbits of $
T$  are fixed points.

A number of issues arise in examining the behavior of the orbits of a given
dynamical system $(\Sigma^n, X)$.  Of primary interest here is whether or not
$(\Sigma^n, X)$ contains any 
periodic orbits ({\em i.e.},  orbits which pass
repeatedly through the same point).
We shall say that a periodic orbit $\lambda$ is an {\em attractor}
if all the nearby orbits  approach
it, and a {\it repeller} if they all move away.
(In general, of course, a periodic orbit is neither a repeller nor an
attractor.)

We wish now to briefly describe some specific examples of dynamical systems
which we will find useful in our discussion of the dynamics of Cauchy horizons:

\noindent{\bf Example 1: } Let $\Sigma^2$ be any two-dimensional
manifold, and let $\psi$ be any diffeomorphism from $\Sigma^2$ to
itself.  Let us recall the {\em suspension} construction \cite{Smale}
of a three-dimensional dynamical system which has global transverse
section $\Sigma^2$ and has Poincar\'e map $\psi$: For the manifold
$\Sigma^3$ of this dynamical system, one chooses the twisted product
$\Sigma^2 \times_\psi S^1$, which is defined by quotienting $\Sigma^2
\times \R$ by the map
\begin{eqnarray*}
  \Ximap: \Sigma^2 \times \R &\to & \Sigma^2 \times \R\ , \\ (p, s) &
  \mapsto& (\psi^{-1} ( p), s + 1)\ .
\end{eqnarray*}
So $\Sigma^3 = \Sigma ^2 \times_\psi S^1\equiv\{\Sigma^2 \times \R\}
/\Ximap$.  Then for the vector field $X$ of the dynamical system, one
chooses $X = \rho_*({\partial/\partial s})$, where $\rho$ is the
natural projection map $\rho:\Sigma^2 \times \R \to \Sigma^2
\times_\psi S^1$ associated with the definition of the twisted product
$\Sigma^2 \times_\psi S^1$, and $\partial/\partial s$ is the vector
field tangent to the $\R$ factor of $\Sigma^2\times \R$.  One easily
verifies that $(\Sigma^3, X) = (\Sigma^2 \times_\psi S_1 ,
\rho_*({\partial\over \partial s}))$ has global transverse sections
diffeomorphic to $\Sigma^2$, and that $\psi$ is the corresponding
Poincar\'e map.

Now let $L$ be any 2 $\times$ 2 matrix with integer entries, unit
determinant, and eigenvalues with nonunit absolute value --- {\em
  e.g.}, $L = \left(\matrix{ 2&3\cr1&2\cr}\right)$. Based on the
lattice quotient definition of the two-torus, any such matrix $L$
defines a diffeomorphism $\psi_L:T^2 \to T^2$ of the two-torus to
itself in a standard way.  Then, using the suspension construction
described above, we obtain for any such $L$ a corresponding dynamical
system $(\Sigma^3_L, X_L)$ with $\psi_L$ for its Poincar\'e map.

One verifies ({\em cf.\ e.g.\/} \cite{PdM}[pp.156--159]) that for any
choice of $L$, the dynamical system $(\Sigma^3_L, X_L)$ on the compact
manifold $\Sigma^3_L$ (with nowhere vanishing generator $X_L$) has the
following properties:
\begin{enumerate}
\item $X_L$ has a countable infinity of periodic orbits. The set of
  all points $p \in\Sigma^3_L$ which lie on periodic orbits of $X_L$
  is dense in $\Sigma^3_L$.
\item There are no attracting or repelling periodic orbits.
\item $(\Sigma^3_L, X_L)$ is {\em ergodic} ({\em cf.\ e.g.\/}
  \cite{Sinai}[Ex.5,p.19]).
\item $(\Sigma^3_L, X_L)$ is {\em structurally stable}, so that all
  the
properties here 
are preserved under all sufficiently small $C^1$ perturbations of the
vector field $X_L$.  \een The dynamical systems $(\Sigma^3_L, X_L)$
will be useful for building (stable) families of spacetimes which have
compact Cauchy horizons with nonfountain--like generator dynamics
({\em cf.\/} Section \ref{Section 3}).

\noindent{\bf Example 2:} Consider\footnote{We are grateful to C.\
  Robinson for pointing out this example to us.} a so-called DA
diffeomorphism $\psi_{\DA }: T^2 \to T^2$, as defined by Smale
\cite{Smale} ({\em cf.\/} also \cite{PdM}).  We do not wish to
describe $\psi_{\DA }$ in detail; however, we wish to note the
following. Let $(\Sigma^3_{\DA }, X_{\DA })$ be obtained by suspension
of $\psi_{\DA }$. Then the dynamical system $(\Sigma^3_{\DA } , X_{\DA
  })$ (with nowhere vanishing generator $X_{\DA }$) exhibits the
following properties \cite{PdM}[pp.165--169]: \ben
\item There is one repelling orbit $\Gamma$; there are no attracting
  orbits.
\item There is a {\em non--periodic} attracting set $\Lambda$ ({\em
    ``strange attractor"}), which is locally the product of $\R $ with
  a Cantor set. Almost every orbit asymptotically approaches
  $\Lambda$, when followed to the future.  $\Lambda$ contains a
  countable infinity of periodic orbits (none of which are attractors
  or repellers).
\item The existence and properties of the attracting set $\Lambda$
  above are preserved under all sufficiently small smooth
  perturbations of the vector field
$X_{\DA }$. 
\item There exists a neighborhood $\cV$ of the repelling orbit
  $\Gamma$ such that an {\em arbitrary} perturbation of $X_{\DA }$
  supported in $\cV$ will {\em not} affect the existence and the
  ``chaotic" character of the attracting set $\Lambda$. (Such a
  perturbation might lead to a different basin of attraction of
  $\Lambda$. The new basin of attraction will nevertheless still have
  nonzero measure.)  \een We will use Example 2 (and some cutting and
  pasting) to build spacetimes containing compactly generated,
  noncompact, ``asymptoticaly flat'' Cauchy horizons with
  nonfountain--like generator dynamics ({\em cf.\/} Section
  \ref{Section 4}).

\section{Compact Cauchy horizons with nonfountain--like dynamics}
\label{Section 3}

In this section we shall show that there exist smooth compact Cauchy
horizons with no attracting periodic orbits. [Since a ``fountain'', as
defined in the Introduction, is precisely an attracting periodic
orbit, the existence of spacetimes with Cauchy horizons with
nonfountain--like behaviour immediately follows.]  We have the
following:
\begin{Proposition}
\label{T4.1}
Let $(\Sigma^3, X)$ be any dynamical system with $\Sigma^3$ compact
and X nowhere vanishing.  There exists a spacetime $(M^4, g)$ (not
necessarily satisfying any field equations and/or energy conditions)
containing a Cauchy horizon ${\cal H}$ which is diffeomorphic to
$\Sigma^3$, and such that the generators of ${\cal H}$
are tangent to the 
orbits of X.
\end{Proposition}

We divide the proof into two main steps, the first of which involves
proving the following Lemma:

\begin{Lemma}
\label{L4.1}
Let $(\Sigma^3, X)$ be any dynamical system with $\Sigma^3$ compact
and X nowhere vanishing.  Consider a spacetime $(M^4, g)$ with $M^4 =
\Sigma^3 \times (-\mu, \mu)$ for some $\mu > 0$, and let $Z$ be a
vector field on $M^4$ such that $Z |_{\Sigma^3 \times \{0\}} = X$.
Suppose moreover that the following hold: \ben
\item
\label{(d)}
$g(Z, Z)\Big|_{\Sigma^3 \times \{0\}} = 0$,
\item
\label{(e)}
$g^{-1}(dt, dt) <0$ for all $t < 0$, where $t$ parametrizes the
interval $(-\mu, \mu)$.  \een { Then}: \ben
\item $(\tilde M ^4, \tilde {g}) \equiv(\Sigma^3\times(-\mu, 0),
  g|_{\tilde M ^4})$ is globally hyperbolic,
\item $\Sigma^3 \times \{0\}$ is a future Cauchy horizon for $(\tilde
  M ^4, \tilde {g})$ in $(M^4, g)$, and
\item $Z|_{\Sigma^3 \times \{0\}} = X$ is tangent to the null
  generators of that Cauchy horizon.  \een
\end{Lemma}

\noindent {\bf Proof of Lemma:} It follows from hypothesis \ref{(e)}
of this Lemma that the function
\begin{eqnarray*}
  T:M^4 =& \Sigma^3 \times(-\mu, \mu)& \to \R \\ &(p, t) &\mapsto t
\end{eqnarray*}
is a time function on $\tilde M ^4$.  Hence we know from Theorem 8.2.2
in Wald \cite{Wald} that the spacetime $(\tilde M ^4, \tilde {g})$ as
defined above is stably causal, and further ({\em cf.\/} the Corollary
on p. 199 of \cite{Wald}) that it is strongly causal.  Now to show
that $(\tilde M ^4, \tilde g)$ is globally hyperbolic, it is
sufficient ({\em cf.\/} p. 206 of Hawking and Ellis \cite{HE}) to
verify that in addition one has $J^+(p)\cap J^-(q)$ compact for every
$p, q \in \tilde M ^4$, where $J^+(p)$ is the closure of the future of
$p$ in $\tilde M ^4$, and $J^-(q)$ is the closure of the past of $q$
in $\tilde M ^4$.  But $J^+(p)\cap J^-(q)$ is certainly closed and it
is also the subset of a compact region $\Sigma^3 \times [T(p), T(q)]
\subset \tilde M ^4$.  Hence $J^+(p)\cap J^-(q)$ is compact, and it
follows that $(\tilde M ^4, \tilde {g})$ is globally
hyperbolic\footnote{Note that this argument shows, that a spatially
  compact stably causal spacetime is necessarily globally
  hyperbolic.}.

Now $Z|_{{\cal H} }$ is nowhere vanishing,  tangent to ${\cal H}$ and null. It
follows that the integral curves of $Z|_{{\cal H} }$ are causal
curves which never leave $\cal {H}$. Hence no subset of $M^4$
containing  $\tilde M^4$ and larger
than $\tilde M^4$ can be globally hyperbolic, and consequently
${\cal H}$ is a Cauchy horizon. By \cite{HE} there is a
unique null direction tangent to each point of a smooth Cauchy
horizon, with the
null generators being tangent to this direction, so it must be that
 $Z|_{{\cal H} }$ is tangent to the null generators at each point of ${\cal
H}$.
\hfill $\Box$

\noindent {\bf Proof of Proposition \ref{T4.1}:} By Lemma \ref{L4.1}
all we need to do now is show that for any dynamical system
$(\Sigma^3, X)$ with $\Sigma^3$ compact and $X$ nonvanishing, we can
always find a spacetime $(M^4, g)$ which satisfies the hypotheses of
the Lemma.  So let $\mu$ be any positive real number, let $t$
parametrize the interval $(-\mu, \mu)$ and set $M^4 = \Sigma^3\times
(-\mu, \mu)$.  The vector field $X$, defined in an obvious way on
$\Sigma^3\times \{0\}$, may now be Lie--dragged along the flow of
$\partial/ \partial t$ to define the vector field $Z$ on $M^4$.  By
construction we have $Z|_{\Sigma^3 \times \{0\}} = X$.  Note that this
construction also guarantees that $dt(Z) = 0$.  To construct the
appropriate spacetime metric, we first arbitrarily choose the
following three fields: \ben
\item \label{i} Let $\phi: (-\mu, \mu) \to \R $ be any monotonically
  decreasing function such that $\phi (0) = 0$.  From $\phi$, we
  construct $\chi: M^4 \to \R $ by setting $\chi(p, t) = \phi(t)$.
\item Let $\beta$ be any one-form on $M^4$ such that $\beta(Z) = 1$
  and $\beta ({\partial/\partial t}) = 0$.
\item Let $\gamma$ be any Riemannian metric on $\Sigma^3$; for $V,W\in
  T\Sigma$ set
$$
\tilde \gamma(V,W)\equiv\gamma(V,W) -{1\over2}\{\gamma(X,
V)\beta(W)+\gamma(X, W)\beta(V)\}\ ,
$$
$$
\tilde \gamma(\frac{\prt}{\prt t},\frac{\prt}{\prt t})\equiv
\tilde \gamma(\frac{\prt}{\prt t},W)\equiv 0
$$ (note that $\tilde \gamma(X, \cdot) = 0)$; and finally define $\nu$
as a symmetric $0\choose 2$ tensor field on $M^4$ by dragging $\tilde
\gamma$ along $\partial \over \partial t$. [Here $\beta$ has been
identified with a form on $\Sigma$ in the obvious way.]  Note that it
follows from this definition that $\nu(Z, Z) = 0$ and $\nu({\partial/
  \partial t}, {\partial/ \partial t}) = 0$.  \een Using $\chi$, $
\beta$, and $\nu$, we define \be
\label{metric}
g = \chi \beta \otimes \beta + dt \otimes \beta + \beta\otimes dt
+\nu\ .  \ee We verify immediately from the properties of $\chi,
\beta$, and $\nu$ and from the definition of $Z$ that $g(Z, Z) = \chi$
everywhere on $M$, and in particular $g(Z, Z)|_{\Sigma^3 \times \{0\}}
= 0$, so hypothesis \ref{(d)} of Lemma \ref{L4.1} is satisfied.  To
verify hypothesis \ref{(e)} it is useful to set up local coordinates
$(x, y, z, t)$ such that $\beta = dz$, $Z = {\partial/\partial z}$; it
follows
that $\nu = \nu_{xx} dx^2 + 2\nu_{xy}dx dy 
+ \nu_{yy}dy$.  Then the components of the metric $g$ take the matrix
form
$$
g_{\alpha\beta} = \left(\matrix{ \nu_{xx} & \nu_{xy}&0 & 0\cr
\nu_{yx}&\nu_{yy} & 0 & 0 \cr
0 & 0 & \chi & 1 
\cr
0 & 0 & 1 
& 0 \cr}\right)\ .
$$ {}From this matrix representation, we see that $g$ is indeed a
Lorentz metric (nondegenerate, signature $+++-)$ and we calculate the
matrix representation of the inverse metric:
$$
(g^{-1})^{\alpha\beta} = \left(\matrix{{1\over \det\nu }  \nu_{yy} & {-1\over
\det\nu}  \nu_{xy}& 0 & 0 \cr
 {-1\over \det\nu }  \nu_{yx} & {1\over \det\nu} \nu_{xx} &0 &0\cr
0&0&0&1\cr
0&0&1&-\chi\cr}\right)\ .
$$
We see that $g^{-1}(dt, dt) = -\chi$, which implies that $g^{-1}(dt,
dt) < 0$ for $t < 0$, as required by hypothesis \ref{(e)} of Lemma
\ref{L4.1}.  So the spacetime $(M^4, g)$ which we have constructed
(from the dynamical system $(\Sigma^3, X))$ satisfies all of the
hypotheses of the Lemma, thus completing the proof of the
Proposition.\hfill $\Box$

Using the example dynamical systems from Section \ref{Section 2},
together with this Proposition (and some of the constructions outlined
in its proof), we can easily construct a large numbers of spacetimes
containing compact Cauchy horizons with nonfountain--like dynamics.
Here the only essential restriction is, that the vector field $X$
generating the dynamical system be nowhere vanishing --- this excludes
examples like {\em e.g.\ } (a compactified version of) the Lorenz
attractor \cite{Sparrow} or of the geometric model thereof \cite{GW},
but clearly allows for interesting dynamics.  Models with {\em e.g.\/}
``horseshoes'' can be constructed on $S^2 \times S^1$ using the
periodically perturbed nonlinear pendulum equation or the periodically
perturbed Duffing equation.

We wish to stress that these examples can be constructed in such a way
that the nontrivial properties of the dynamics are {\em stable under
  small smooth variations of the metric.} For example, let
$(\Sigma^3,X)$ be the Anosov flow discussed in Example 1 of the
previous Section. The metric $g$ constructed in the proof of
Proposition \ref{T4.1} can be chosen to satisfy the stability
criterion of \cite{ChI}, so that small smooth variations of the metric
will lead to small $C^k$ variations\footnote{$k$ here may be made
  arbitrarily large (but probably {\em not} $k=\infty$) by
  appropriately choosing $g$.} of the Cauchy horizon. This in turn
will lead to a small $C^{k-1}$ variation of the field of null tangents
to the generators, and the stability of the resulting dynamical
systems follows from stability of Anosov flows.

Note that all examples discussed so far have $\Sigma^3$ defined as a
twisted product of a two--dimensional manifold with the circle.  Do
all compact Cauchy horizons with nonfountain--like behavior have this
sort of topology?  Certainly not.  In the next Section we shall see
how to construct Cauchy horizons with interesting dynamical behaviour
of the generators by using the connected sum operation.  It would be
of some interest to find out whether or not there are spacetimes with
nonfountain--like behavior in a Cauchy horizon of arbitrary (compact,
three-dimensional) topology.

\section{Noncompact compactly generated horizons}
\label{Section 4}
It is relatively easy to construct a spacetime $(M^4,g)$ which has a
compactly-generated but noncompact Cauchy horizon with
nonfountain--like dynamics. First, one chooses a compact dynamical
system $(\Sigma^3,X)$ which has nonfountain--like dynamics and also
has a repelling periodic orbit: the DA system as discussed in example
2 will do.  Then, one uses Proposition \ref{T4.1} to construct a
spacetime containing a Cauchy horizon diffeomorphic to $\Sigma^3$ with
generators matching the orbits of $X$. Finally one removes the
repelling orbit from the horizon in the spacetime. The Cauchy horizon
$\cH$ of the resulting spacetime is clearly not compact. On the other
hand, one verifies that $\cH$ is compactly-generated by noting that if
one defines the set $\cK=\cH\setminus \tilde S$, where $\tilde S$ is a
small open thickening of the removed orbit, then since the removed
orbit was repelling, all past-directed null generators of $\cH$ enter
and remain in $\cK$, which is compact.  [Note that this example shows
that the inequality $f \ge 0$, which according to \cite{Hawking} holds
for any periodic generator of a Cauchy horizon, is not correct.]

The above example is rather artificial, and it is natural to enquire
about the existence of {\em smooth} compactly generated horizons in
asymptotically flat spacetimes.  By way of example, consider $M =
\R^4$ with a metric $g$ which is the standard Minkowski metric outside
of a compact set $\cC$. Let us moreover assume that there exist
periodic time-like curves in $\cC$. [An explicit example of such a
spacetime can be found in \cite{Hawking}.]  $M$ will have a Cauchy
horizon ${\cal H}^+$, which is the boundary of the domain of
dependence of any standard $t = const$ plane lying in $ M\setminus
J^+(\cC)$.  Now ${\cal H}^+$ can be ``sandwiched" between $\partial
J^+(p)$ and $\partial J^+(q)$, where $p, q$ are any two points such
that $\cC \subset J^+(p)$, and $\cC \cap J^+(q) = \emptyset$; by
``sandwiched" here we mean that ${\cal H}^+\subset \{I^-(\prt
J^+(q))\cap I^+(\prt J^+(p))\} $.  It is then easily seen that for all
$R \in \R$ large enough the world tube ${\cal T} = \{(t, \vec x): t
\in\R, |\vec x| = R\}$ intersects each of the generators of ${\cal
  H}^+$ transversally.

Based on this example, we shall say that a compactly generated Cauchy
horizon $\cH$ in an orientable and time--orientable spacetime $(M,g)$
is {\em of asymptotically flat type} if the boundary set $\partial
(\cK\cap \cH)$ consists of a finite number $I$ of spheres $S_i$, with
each generator of $\cH\setminus \cK$ intersecting one of the $S_i$'s
transversally. Here $\cK$ is one of the compact sets in $M$ which
characterizes $\cH$ as compactly generated. It is easy to convince
oneself that the behaviour described in this definition should occur,
{\em e.g.\/}, for compactly generated Cauchy horizons in spacetimes
which admit a sufficiently regular compactification in lightlike
directions (the number $I$ above corresponds then to the number of
connected components of Scri).

We would like to find spacetimes with compactly-generated,
asymptotically flat type Cauchy horizons with nonfountain-like
dynamics. The following result is an obstacle to the
construction of such spacetimes:

\begin{Proposition}
\label{P3}
Let ${\cal H}^+$ be a compactly generated future Cauchy horizon of
asymptotically flat type.  Then the field $X$ of directions tangent to
the generators of ${\cal H}^+$ {\em cannot} be continuous.
\end{Proposition}

\noindent {\bf Proof}: Suppose that the field $X$ is continuous. Now
consider the compact manifold $\hat{\cH}$ constructed by adding a
point $p^i_\infty$ to each of the ``asymptotic ends" $S_i\times
\R\subset \cH^+$. We can deform the field of generators on each of the
ends $S_i\times \R$ to obtain a continuous vector field $\hat X$ on
$\hat{\cH}$ which is nowhere vanishing except at the points
$p^i_\infty$. At each of those points the index of $\hat X$ will be
equal to $+1$; consequently the index of $\hat X$ will be equal to
$I\ne 0$. Note that $\hat{\cH}$ is orientable because $(M,g)$ has been
assumed to be time--orientable and orientable. This, however,
contradicts the fact that the index of a continuous vector field on a
compact, three--dimensional, orientable manifold vanishes. \hfill
$\Box$

This result makes it difficult to systematically study the dynamics of
the null generators in spacetimes containing compactly generated
Cauchy horizons of asymptotically flat type. In particular, the
construction carried out in Proposition \ref{T4.1} encounters various
obstacles.  However, as it has been suggested
\cite{Thorne,FMNEGT,Hawking} that there exist compactly generated
Cauchy horizons ${\cal H}$ of asymptotically flat type which are
smooth on an open dense set $\cU$ (the complement of which has zero
measure), we believe the following result should be of
interest:\footnote{Here we define $\prt^+\cD(\Sigma)=
  \overline{\cD^+(\Sigma)}\setminus(\cD^+(\Sigma)\cup\Sigma)$,
  similarly for $\prt^-\cD(\Sigma)$.  We use the convention in which
  the domains of dependence are {\em open} sets; in particular they do
  {\em not} include the Cauchy horizons.}

\begin{Proposition}
\label{p4}
Let ${\cal H}^+=\prt^+\cD(\Sigma)$ be a future Cauchy horizon in a
space-time $(M,g)$. Suppose that there exists an open subset $\cU$ of
${\cal H}^+$ such that $\cU$ is a smooth submanifold of $M$.  Suppose
moreover that there exists a smooth time function $\tau $ on
$\cD(\Sigma)$ such that $\lim_{\cD(\Sigma)\ni p\to\cU}\nabla\tau(p)$
exists, and is a smooth, nowhere vanishing vector field on $\cU$.
Then there exists a space-time $(M' , g' )$ with a future Cauchy
horizon ${\cal H}'$ diffeomorphic to {\em $\cH^+\#\Sigma^3_{\DA}$}
(where {\em $\Sigma^3_{\DA}$} is the manifold discussed in Example 2,
Section \ref{Section 2}), and with non--trivial long--time dynamics of
the generators of ${\cal H}' $.  [Here $\#$ denotes the connected
sum.]  Moreover, ${\cal H}' $ will share certain overall properties of
${\cal H}^+$; in particular if ${\cal H}^+$ is compact, or compactly
generated, or of asymptotically flat-type, then the same will be true
of ${\cal H}' $.
\end{Proposition}

\noindent {\bf Remarks:} We believe that the inclusion in Proposition
\ref{p4} of the hypothesis that the function $\tau$ exists should be
unnecessary, for the following reasons:
\begin{enumerate}
\item We consider it likely that the remaining conditions of
  Proposition \ref{p4} are sufficient to guarantee that such a
  function can be constructed.
\item We have written the proof below in such a way that the existence
  of the function $\tau$ is essentially used in one place only. We
  believe that it should be possible to replace that step of the
  argument by one which does {\em not} require the existence of the
  function $\tau$.
\end{enumerate}

\noindent {\bf Proof:} Let $\Sigma$ be a partial Cauchy surface in
$(M,g)$ such that $\cH^+=\prt^+\cD(\Sigma)$.  Replacing $M$ by a
subset thereof if necessary we may assume that
$\cH^-\equiv\prt^-\cD(\Sigma)= \emptyset$. Let $\cU$ be a smooth
subset of ${\cal H}^+ $. Passing to a subset of $\cU$ if necessary we
may without loss of generality assume that: 1) the closure $\bar
{\cU}$ of $\cU$ is compact, and 2) that the generators of ${\cal H}^+$
have a cross-section $S$ in $\cU$, and 3) that $\cU \approx S \times
(-1,1)$, with $S$ being a smooth two-dimensional embedded submanifold.
We claim that we can find a defining function $\varphi$ for $\cU$,
defined on a conditionally compact neighborhood $\cO$ of $\cU $, such
that $\varphi|_{{\cal O}\cap \cD^+(\Sigma)}$ is a time function.
[Recall that $\varphi : {\cal O} \rightarrow \R$ is a defining
function for $\cU $ if $d \varphi$ is nowhere vanishing on $\cU$ and
if we have $p \in \cU\cap\cO \Leftrightarrow \varphi (p) = 0$.]  If we
have a time function $\tau$, as assumed in the hypotheses of
Proposition \ref{p4}, we set $\varphi=\tau|_{{\cal O}\cap
  \cD^+(\Sigma)}$, and we are done.

[Had we not made the assumption of the existence of $\tau$, the
existence of $\varphi$ could be established as follows: Let $\psi$ be
any defining function for $\cU $ defined on some neighborhood ${\cal
  O}$, and let $X$ be any future--directed timelike vector field on
${\cal O}$.  If \be\label{timeineq} [X^\mu\nabla_\mu(\nabla^\nu
\psi\nabla_\nu \psi)]\Big|_\cU > 0, \ee then passing to a subset of
${\cal O}$ if necessary we shall have $\nabla^\nu \psi \nabla_\nu
\psi|_{\cD^+(\Sigma)\cap {\cal O}} < 0$, and then setting $\varphi =
\psi$ we are done.  If \eq{timeineq} does not hold, consider any
smooth function $\alpha$ on ${\cal O}$; we have \be
X^\mu\nabla_\mu\left(\nabla^\nu(\alpha\psi)\nabla_\nu
(\alpha\psi)\right)\Big|_\cU =
\alpha^2[X^\mu\nabla_\mu(\nabla^\nu\psi\nabla_\nu\psi) +
X^\nu\nabla_\nu\psi\nabla^\mu \psi\nabla_\mu(\log^2\alpha))]\ .
\label{C.1} \ee Note that $X^\nu \nabla_\nu \psi$ is nowhere
vanishing on $\cU $, as $X^\nu $ is time-like and $\nabla^\nu \psi$ is
null.  Let $\hat\alpha : \cU \rightarrow \R$ be any strictly positive
solution of the equation
$$[\nabla^\mu\psi\nabla_\mu( \log^2\hat\alpha)] =
(X^\mu\nabla_\mu\psi)^{-1} [1 -
X^\mu\nabla_\mu(\nabla^\nu\psi\nabla_\nu\psi)]\Big|_\cU ,$$ and let
$\alpha$ be any strictly positive extension of $\hat\alpha$ to $\cO $.
Setting $\varphi=\alpha\psi$ the desired defining function then
follows.  Passing to a subset of ${\cal O}$ we may moreover assume
that $d\varphi$ is nowhere vanishing on ${\cal O}$.  Changing
$\varphi$ to $-\varphi$ if necessary we may suppose that $\nabla^\nu
\varphi$ is past--directed on $\cD^+(\Sigma)\cap {\cal O}$.]

Let ${\cal B}_{4\rho}\subset \cU $ be a closed coordinate ball of
radius $4\rho$ covered by coordinates $x^i$, with the $x^i$'s chosen
so that $g^{\mu\nu}\varphi_{,\mu}|_\cH={\prt\over\prt x^3}$, and with
${\cal B}_{4\rho}$ compact in $\cU $.  If we choose a timelike future
directed vector field $T$ on ${\cal O}$, then by dragging the
coordinates $x^i$ along the integral curves of
$T$ 
we obtain a coordinate system $(x^0,x^i)=(\varphi,x^i)$ on a compact
set $[-\delta, \delta ] \times {\cal B}_{4\rho}$, for some $\delta >
0$.  Since $\varphi$ is a time function on $[-\delta, 0 ) \times {\cal
  B}_{4\rho}$, the sets $\{s\} \times {\cal B}_{4\rho}$ are spacelike
for $s \in [-\delta, 0 )$. In this coordinate system the metric takes
the form \be
\label{firstmetric}
g_{\mu\nu} dx^\mu dx^\nu= 2g_{30}dx^0dx^3 + g_{00}(dx^0)^2+
2g_{0A}dx^0dx^A + g_{AB}dx^Adx^B+O(\varphi)\ , \ee where the labels
$A,B$ run over $1,2$, and where $O(\varphi)$ indicates terms which
vanish at least as fast as $|\varphi|$ for small values of $|\varphi|
$.  Consequently, \be
\label{det}
\det g = -(g_{30})^2 \det(g_{AB}) +O(\varphi)\ , \ee from which it
follows that if $\delta$ is sufficiently small, then $g_{30}$ does not
change sign. {}From the above construction, it follows that in fact
$g_{30}$ is positive.

Let $t$ be any strictly negative time function on $\cD(\Sigma)$ such
that 1) the level sets of $t$ are Cauchy surfaces for $\cD(\Sigma)$,
and 2) $t(p) \rightarrow 0$ as $p \rightarrow \partial^+ \cD(\Sigma)$.
Then set
$$
\epsilon = \inf  -t(p)\  |\    p \in \{-\delta\} \times  {\cal B}_{2\rho}.
$$
Now, consider the spacetime  region $(M_1,g^1)$ with
\be
\label{m1}
M_1=M\setminus \Big(
J^-(\{-\delta\}\times {\cal B}_{2\rho})\cup[\{\delta\} \times {\cal
  B}_{2\rho}]\Big), \qquad g^1=g\Big|_{M_1}\ .  \ee Clearly, $\Sigma=
\{t = -\epsilon/2\}$ is a partial Cauchy surface in $M_1$ such that
$\partial^+\cD(\Sigma;M_1)={\cal H}^+$ is a future Cauchy horizon for
$\Sigma$. Here we use the notation $\cD(\Sigma;M_1)$ for the domain of
dependence of $\Sigma$ in $(M_1,g^1)$; we shall use a similar
convention for $J^\pm$, etc.

Let $\Sigma^3_{\DA }$ be the manifold discussed in Example 2, Section
\ref{Section 2}.  Let $\Gamma$ be the repelling orbit and $\cV$ be the
designated neighborhood of $\Gamma$ as discussed in that Example, and
let ${\cal B}_{4\rho}^1{\subset \cV }$ be a closed coordinate ball
covered by coordinates $y^i$.  Finally let $\psi$ be the inversion
map:
$$ \cU\supset\wnetrze{\cB}_{2\rho}\setminus{\cB}_{\rho/2} \ni
x^i\stackrel{\psi}{\rightarrow} y^i=-{x^i\over r(x)^2} \in
\wnetrze{\cB}^1_{2\rho}\setminus{\cB}^1_{\rho/2}\subset
\Sigma^3_{\DA}\ ,
$$
where $r(x)=\sqrt{\sum (x^i)^2}$. It is easily shown that one can
find a nowhere vanishing vector field $X$ on $ \Sigma^3_{\DA
  }\setminus{\cal B}_\rho^1 $ such that
$$
(\psi^{-1})_* X\Big|_{{\cal B}_{2\rho}\setminus \wnetrze{{\cal
      B}}_{\rho}} = \nabla \varphi\ ,
$$
and
$$
X\Big|_{\Sigma^3_{\DA }\setminus{\cal B}_{3\rho/2}^1}=X_{\DA}\ ,
$$ where $X_{\DA}$ is the generator of the DA flow discussed in
Section 2, and where we have used ``hats" to denote the interior of a
set: $\wnetrze{{\cal B}}_{\rho} = \mbox{int}\, {\cal B}_{\rho}$, etc.
Now let $g^{\DA}$ be any Lorentzian metric constructed on $M_{\DA }
\equiv (-\delta, \delta ) \times \{\Sigma^3_{\DA }\setminus {\cal
  B}_{\rho/2}^1\}$ as described in the Proof of Proposition
\ref{T4.1}.  On $(-\delta, 0 ) \times \{\Sigma^3_{\DA }\setminus {\cal
  B}_{\rho/2}^1\}\subset M_{\DA }$, we can define a time function
$y^0$ by
$$
y^0(s,p) = s\ .
$$
Based on the map $\Psi$ we define
$$
(-\delta,\delta)\times \{\wnetrze{{\cal B}}_{2\rho}\setminus {{\cal
    B}}_{\rho/2}\}\ni (t,p)\rightarrow \Psi(t,p)=(t,\psi(p))\in
(-\delta,\delta)\times \{\wnetrze{{\cal B}}_{2\rho}^1\setminus {{\cal
    B}}^1_{\rho/2}\} \ .
$$ Then, in local coordinates on $(-\delta,\delta)\times
\{\wnetrze{{\cal B}}_{2\rho}\setminus {{\cal B}}_{\rho/2}\}$ the
metric $\Psi^*g^{\DA}$ takes the form \be
\label{secondmetric}
g^{\DA}_{\mu\nu}dx^\mu dx^\nu=2\beta_\mu dx^0
dx^\mu+g^{\DA}_{ij}dx^idx^j\ , \ee where
$$
\beta_0=0,\qquad \beta_idx^i = \psi^*\beta \ .
$$ On $\wnetrze{{\cal B}}_{2\rho}\setminus {{\cal B}}_{\rho}$ we have
\be
\label{intermed}
\beta_3=\beta_\mu\nabla^\mu\varphi=\langle
\Psi^*\beta,\nabla\varphi\rangle =
\langle\beta,\Psi_*\nabla\varphi\rangle = \langle\beta,X\rangle = 1 \
.  \ee Since the metric $g^{\DA}$ is $y^0$--independent, \eq{intermed}
actually holds on $(-\delta,\delta)\times \{\wnetrze{{\cal
    B}}_{2\rho}\setminus {{\cal B}}_{\rho}\}$.  A similar calculation
shows that \be
\label{intermed2}
g_{3i}^{\DA}=0 \ee on $(-\delta,\delta)\times \{\wnetrze{{\cal
    B}}_{2\rho}\setminus {{\cal B}}_{\rho}\}$.  Now let $\phi\in
C^\infty(\R^4)$ be any non--negative function such that $\phi=0$ in
$\R\times \cB_{\rho}$, and $ \phi=1$ in $\R^4\setminus (
\R\times\cB_{2\rho})$.  On $M_1$ we may define the smooth metric $g^2$
to coincide with $g^1$ on
$M\setminus\{(-\delta,\delta)\times\cB_{2\rho}\} $, and to be given by
\be
\label{thirdmetric}
g^2_{\mu\nu}dx^\mu dx^\nu= \phi g_{\mu\nu}dx^\mu dx^\nu + (1-\phi)
g^{\DA}_{\mu\nu}dx^\mu dx^\nu \ee on
$(-\delta,\delta)\times\{\cB_{2\rho}\setminus\cB_{\rho/2}\}$ (it is
easily seen from eqs.\ \eq{firstmetric},
\eq{secondmetric}--\eq{thirdmetric} and from eq.\ \eq{det} with $g^2$
substituted for $g$, that \eq{thirdmetric} indeed defines a Lorentzian
metric). Note that $\varphi=x^0$ is still a time function for this new
metric in $(-\delta,0)\times\{\cB_{2\rho}\setminus\cB_{\rho/2}\}$.

The desired spacetime $M'$ will now be obtained by gluing together
$(M_1,g^2)$ and $(M_{\DA },g^{\DA})$:
Specifically, we choose
\be
\label{MS}
M' = \Big[\{M_1\setminus [ (-\delta, \delta ) \times \wnetrze{{\cal
    B}}_{\rho/2}]\} \sqcup M_{\DA} \Big]/\Psi\ .  \ee Since the
metrics $g^2$ and $\Psi^*g^{\DA}$ coincide on
$\cB_\rho\setminus\cB_{\rho/2}$, a metric $g'$ can be defined on $M'$
in the obvious way. There is a natural identification between points
in ${M_1\setminus [ (-\delta, \delta ) \times \wnetrze{{\cal
      B}}_{\rho/2}]}$ and an appropriate subset of $M'$, and similarly
for points in $M_{\DA}$, with another subset of $M'$.  We shall use
this identification without mentioning it explicitly in what follows.
Let us note that the function $\varphi'$, defined as
$$ \varphi'(p)=\cases{\varphi(p)\ (=x^0(p)), & $p\in{ (-\delta, 0]
    \times[ \cB_{4\rho}\setminus\wnetrze{{\cal B}}_{\rho/2}]}$, \cr
  y^0(p),&$ p\in (-\delta, 0 ] \times[ \Sigma^3_{\DA} \setminus
  \wnetrze{{\cal B}}_{\rho/2}^1]$,}
$$ is a smooth time function on the interior of the set on which it
has been defined.

We now claim that  the submanifold $\mzero $ of $M'$ defined by
$$ \mzero = \Big[\{\cD(\Sigma;M_1)\setminus [ (-\delta, 0 ) \times
\wnetrze{{\cal B}}_{\rho/2}]\} \sqcup \{ (-\delta, 0 ) \times[
\Sigma^3_{\DA} \setminus \wnetrze{{\cal B}}_{\rho/2}^1] \}\Big]/\Psi\
,$$ with the metric obtained from $g'$ by restriction, is globally
hyperbolic.  First, we wish to show that for all $p,q\in \mzero $, the
set $J^+(p;\mzero )\cap J^-(q;\mzero )$ is compact. To do this it is
convenient to consider various cases, according to whether $p\in
\cD(\Sigma;M_1)\setminus [ (-\delta, 0 ) \times \wnetrze{{\cal
    B}}_{\rho}]$, $p\in (-\delta, 0 )
\times [\wnetrze{{\cal B}}_{\rho}\setminus 
{{\cal     B}}_{\rho/2}] $, or
$p\in (-\delta, 0 ) \times[
  \Sigma^3_{\DA} \setminus \wnetrze{{\cal B}}_{\rho/2}^1]$,
similarly for $q$.  Suppose, {\em
  e.g.}, that $p,q\in M_1\setminus [ (-\delta, 0 ) \times
\wnetrze{{\cal B}}_{\rho}]$. We define
$$
K = J^+(p;M_1)\cap  J^-(q;M_1)\cap [ (-\delta, 0 ) \times\prt
{{\cal B}}_{\rho}]\ .
$$
$K$ is easily seen to be compact by global hyperbolicity of
$(\cD(\Sigma;M_1),\gone )$. If $K=\emptyset$ we have $J^+(p;\mzero
)\cap J^-(q;\mzero  )= J^+(p;M_1)\cap J^-(q;M_1 )$ and we are done;
otherwise we have
$$
-\delta < s_-=\inf \varphi(p) < s_+ = \sup\varphi(p) < 0 \,
$$ where the $\sup$ and the $\inf$ are taken over $p\in K$. Since we
have a time function $\varphi'$ on $\mzero \setminus \{ M_1\setminus [
(-\delta, 0 ) \times \wnetrze{{\cal B}}_{\rho}]\}$ which agrees with
$\varphi$ on $K$, it follows that $J^+(p;M')\cap J^-(q;M')$ can be
covered by the compact sets $ J^+(p;M_1)\cap J^-(q;M_1)$, $[ -s_-, s_+
]\times[ {{\cal B}}_{\rho}\setminus \wnetrze{{\cal B}}_{\rho/2}]$ and
$[ -s_-, s_+ ] \times [\Sigma_{\DA}\setminus \wnetrze{{\cal
    B}}_{\rho/2}^1]$.  Using similar arguments one shows compactness
of $J^+(p;\mzero )\cap J^-(q;\mzero )$ for the remaining cases.

To prove  strong causality of $\mzero  $, we use the existence of the
time
function $\tau$ on $\cD(\Sigma,M)$: Indeed, the function $\tau'$ defined
by:
$$
\tau'(p)=\cases{\tau(p), & $p\in M_1\setminus [ (-\delta, \delta  )
  \times \wnetrze{{\cal     B}}_{\rho/2}]$,
\cr
y^0(p), & $p\in M_{\DA}$, }
$$ is a smooth time function on $\mzero$. This ensures strong
causality of $\mzero$, and global hyperbolicity of $\mzero $ follows.
[Let us emphasize, that this is the only point at which the hypothesis
of existence of $\tau$ is needed\footnote{We believe that $\mzero$ as
  constructed here is strongly causal even without the assumption of
  existence of the function $\tau$; we have, however, not been able to
  prove this assertion.} in our argument.]

We wish to show now that the set ${\cal H}'$, defined as the boundary
of $\mzero $ in $M'$, is a Cauchy horizon.  Note first that the
generators of $\cH^+$ in $\cB_\rho\setminus\wnetrze{\cB}_{\rho/2}$ are
the integral curves of $\nabla\varphi=\nabla'\varphi$, where $\varphi$
is the defining function for $\cH^+$ on $\cU$ defined at the beginning
of this proof, and $\nabla'$ is the gradient with respect to the
metric $g'$. These curves are smoothly continued by the null (with
respect to the metric $g'$) integral curves of $\nabla' \varphi'$.
Consider now any subset $\mhat$ of $M'$ which contains $\mzero $ as a
proper subset. It follows that $\mhat\cap {\cal H}' \ne \emptyset$.
Let $p\in \mhat\cap {\cal H}'$. We claim that there exists a past
directed causal curve $\lambda$ through $p$ which never enters $\mzero
$: If $p\in M_1$, then consider a generator $\Gamma$ of $\cal H^+$
through $p$. If $\Gamma$ never enters $\cB_\rho$ when followed
backwards with respect to the time orientation, then the connected
component of $\Gamma\cap\mhat$ which contains $p $ provides the
desired curve $\lambda$. If $\Gamma$ enters $\cB_\rho$, let $\tilde
\Gamma$ be the segment of $\Gamma$ up to the point $\tilde p$ where it
first enters $\cB_\rho$.  $\tilde\Gamma$ can be smoothly continued at
$\tilde p$ by the integral curve $\Gamma'$ of $\nabla' \varphi'$. If
this curve never exits $\cB_\rho$ through the sphere $\prt\cB_\rho$,
we can set $\lambda$ to be the connected component of
$(\Gamma\cup\Gamma')\cap\mhat$ which contains $p$. It it exits
$\cB_\rho$ through the sphere $\prt\cB_\rho$, then it can be smoothly
continued by a segment $\Gamma''$ of a generator of ${\cal H}^+$. If
$\Gamma''$ never reenters $\cB_\rho$, we define $\lambda$ to be the
connected component of $(\Gamma\cup\Gamma'\cup \Gamma'')\cap\mhat$
that contains $p$. If $\Gamma''$ reenters $\cB_\rho$ when followed
backwards with respect to the time orientation, we continue the
procedure above to eventually obtain an inextendible curve $\lambda$
through $p$. This shows the existence of an inextendible,
past--directed causal curve $\lambda\subset\cH'\cap \mhat$ through all
$p\in \cH'\cap \mhat$.  Hence it follows that
$$
\cH'=\prt^+\cD(\Sigma';M')
$$
for some $\Sigma'\subset M'$.

Clearly we have
$${\cal H}' \approx {\cal H}^+ \# \Sigma^3_{\DA }, $$ where \# denotes
the connected sum. Moreover the generators of ${\cal H}' $ coincide
with \begin{enumerate}
\item those of ${\cal H}^+$ on ${\cal H}^+\setminus {\cB}_{\rho}$, and
  with
\item the integral curves of the suspension of the DA--diffeomorphism
  on $\Sigma^3_{\DA }\setminus {\cal B}_{3\rho/2}^1$.
\end{enumerate}
{}From what has been said in Example 2, Section \ref{Section 2}, it
follows that a ``non-zero measure" set of generators of ${\cal H}' $
will be attracted to a ``strange attractor", when followed backwards
in time.  \hfill $\Box$

We expect that some of the examples constructed as in Proposition
\ref{p4} are stable in the dynamical sense. However, we have no proof
of this assertion. To establish stability one would need to prove that
small smooth variations of the metric lead to small $C^2$ variations
of the horizon on (perhaps an open subset of) $\cU$. Now it is not
difficult to show that, for an appropriately chosen $(M,g)$, the
metric $g'$ on $M'$ can be constructed so that small variations of
$g'$ will indeed lead to small $C^0$ variations of the horizon $\cH'$
({\em cf.  e.g.\/} \cite{BK} for various results of this kind). The
transition from $C^0$ to $C^2$ seems, however, to be a non--trivial
matter.  In particular, we have not been able to generalize the
methods of \cite{ChI} from compact Cauchy horizons with global
cross--sections to noncompact Cauchy horizons.

\section*{Acknowledgements}
\indent We acknowledge a fruitful collaboration with J.\ Friedman at
an early stage of the work on this paper.  Useful discussions with B.\
Birnir, V.\ Moncrief, C.\ Robinson and K.\ Thorne are also
acknowledged. The authors are grateful to the Institute for
Theoretical Physics of the University of California at Santa Barbara
for hospitality and financial support (under NSF grant PHY89--04035)
during part of the work on this paper.  They were also supported in
part by NSF grant PHY90--12301 to University of Oregon. Moreover, J.I.
was supported in part by NSF grant PHY93-8117, while P.T.C.\ was
supported in part by KBN grant \# 2 1047 9101, by NATO and by the
Alexander von Humboldt Foundation.

 \end{document}